\documentclass[aps,pre,groupedaddress,showpacs]{revtex4-1}
\usepackage{amssymb,amsmath,graphicx,dsfont}

\usepackage{color}

\begin{document}


\title{Conditional reversibility in nonequilibrium stochastic systems}
%
\author{Marcus V. S. Bonan\c{c}a}
\email[]{mbonanca@ifi.unicamp.br}
\affiliation{Instituto de F\'isica 'Gleb Wataghin', Universidade Estadual de Campinas, 13083-859, Campinas, S\~{a}o Paulo, Brazil}
%
\author{Christopher Jarzynski}
\email[]{cjarzyns@umd.edu}
\affiliation{Department of Chemistry and Biochemistry and Institute of Physical Sciences and Technology, University of Maryland, College Park, Maryland 20742, USA}

\date{\today}

\begin{abstract}
For discrete-state stochastic systems obeying Markovian dynamics, we establish the counterpart of the {\it conditional reversibility theorem} obtained by Gallavotti for deterministic systems  [Ann. de l'Institut Henri Poincar\' e (A) 70, 429 (1999)].
Our result states that stochastic trajectories conditioned on opposite values of entropy production are related by time reversal, in the long-time limit.
In other words, the probability of observing a particular sequence of events, given a long trajectory with a specified entropy production rate $\sigma$, is the same as the probability of observing the time-reversed sequence of events, given a trajectory conditioned on the opposite entropy production, $-\sigma$, where both trajectories are sampled from the same underlying Markov process.
To obtain our result, we use an equivalence between conditioned (``microcanonical'') and biased (``canonical'') ensembles of nonequilibrium trajectories.
We provide an example to illustrate our findings.
\end{abstract}


\maketitle


\section{Introduction \label{sec:intro}}

In equilibrium statistical mechanics, the equivalence of ensembles in the thermodynamic limit provides a useful tool for analyzing systems that are subject to sharp constraints.
For example, for purposes of calculating averages in physical situations in which the total energy is fixed, the microcanonical ensemble can be replaced by the typically more convenient, fixed-temperature canonical ensemble, provided the temperature is chosen appropriately \cite{reif,touchette2011}.
Alternatively, if we wish to construct an ensemble that describes an equilibrium system conditioned on a specific value of an extensive observable, then we can use Boltzmann-like weights to bias the distribution toward that value; in the thermodynamic limit the conditioned and biased distributions become equivalent.
The mathematical tools for constructing such ensembles have been analyzed rigorously within the theory of large deviations \cite{touchette2009}.

In the last decade these tools have been applied extensively to probability distributions $\mathcal{P}[\mathcal{S}]$ on the space of {\it paths} or trajectories of nonequilibrium systems \cite{evans2004,evans2005,maes2008,evans2010,jack2010,chetrite2013,chetrite2014}.  This approach has been used to study dynamical phase transitions in kinematically constrained models \cite{garrahan2007,vaikuntanathan2014,gingrich2014b}, glass transitions \cite{merolle2005,hedges2009,chandler2010}, quantum systems \cite{ates2012,genway2012,hickey2012}, and efficiency fluctuations in stochastic heat engines \cite{verley2014a,gingrich2014a,verley2014b}. By analogy with the equilibrium case, nonequilibrium path ensembles can be constructed by introducing exponential, Boltzmann-like weights $e^{-\lambda\mathcal{A}[\mathcal{S}]}$ to modify a given probability distribution.  Here the quantity inside the exponent is a \textit{time}-extensive functional of the trajectory, $\mathcal{A}[\mathcal{S}]$, multiplied by a biasing parameter $-\lambda$. The formal analogy with the equilibrium case suggests that such biased ensembles may be equivalent, in the appropriate limit, to nonequilibrium ensembles conditioned on specified values of the path observable $\mathcal{A}$.  This problem has been addressed explicitly by Jack and Sollich \cite{jack2010} for discrete-state systems, and by Chetrite and Touchette \cite{chetrite2013,chetrite2014} for a broad class of stochastic models including diffusive processes.  Using the theory of large deviations, these authors have established the equivalence between conditioned (``microcanonical'') and biased (``canonical'') path space distributions, in the long-time limit, given certain conditions related to the fluctuations of the observable appearing in the exponential weight.  The results contained in Refs.~\cite{jack2010,chetrite2013,chetrite2014} are related to earlier results by Evans~\cite{evans2004,evans2005,evans2010}, obtained within the framework of maximum-entropy inference, as well as to the generalized Onsager-Machlup theory developed by Maes and Neto\v{c}n\'{y}~\cite{maes2008}.

Here we use this equivalence to study nonequilibrium systems that are conditioned on values of entropy production.  For a discrete-state system whose time-averaged rate of entropy production takes a positive value $\bar\sigma$ in the infinite-time limit, Eq.~(\ref{eq:entropy-currentUNB}), we consider the statistical fluctuations in the entropy production rate over {\it finite} intervals of duration $\mathcal{T}$.  Heuristically, it is useful to imagine ``chopping'' an infinite trajectory into segments of duration $\mathcal{T}$, and then segregating these according to the time-averaged entropy production rate during each segment.  Those segments for which the entropy production rate takes on a value $\sigma$ comprise an ensemble that is conditioned on that value.  In the long-$\mathcal{T}$ limit, we find that ensembles conditioned on \textit{opposite} values of entropy production rates are related by time-reversal.  In effect, if we compare two long trajectory segments, conditioned on entropy production rates $\pm\sigma$, then (statistically) one of them will look like a mirror image of the other, in time.  This result is the stochastic counterpart of the {\it conditional reversibility theorem} derived by Gallavotti \cite{gallavotti1999} in the context of deterministic dynamics.

In our presentation we will not aim at full mathematical rigor, nor will we assume that the reader is deeply familiar with large deviation theory.  We will show that our central result, which is the conditional reversibility described above, follows from relatively straightforward manipulations.  In order to keep the presentation self-contained, in Sec. \ref{sec:cond} we will derive results obtained previously in Refs. \cite{jack2010,chetrite2013,chetrite2014}.

The outline of the paper is as follows. In Sec.~\ref{sec:cond} we define a stationary, discrete-state Markov process that violates detailed balance, Eq.~(\ref{cond1}); and following \cite{jack2010,chetrite2013,chetrite2014} we construct the rate matrix for the biased ensemble, Eq.~(\ref{dual6}). In Sec.~\ref{sec:rever} we establish that ensembles biased toward opposite values of entropy production rates $\pm\sigma$ are described by rate matrices that are the dual of one another, Eqs.~(\ref{eq:dualMainResult}), (\ref{eq:entropyTR}), and we use this result to formulate a conditional reversibility theorem for stochastic dynamics, Eq.~(\ref{eq:condRev}).  In Sec.~\ref{sec:example} we present an illustrative example and in Sec.~\ref{sec:conclu} we finish with concluding remarks.

\section{Conditioned and biased ensembles and their dynamics \label{sec:cond}}

We consider a model with $K$ discrete states labelled by $i \in \Lambda \equiv \{1,\ldots,K\}$. The probability $p_{i}(\tau)$ to find the system in state $i$ at time $\tau$ evolves according to the master equation
\begin{equation}
\frac{d p_{i}}{d \tau} = \sum_{j\neq i}^{K} \left( R_{ij}\, p_{j} - R_{ji}\, p_{i} \right) = \sum_{j=1}^{K} R_{ij}\, p_{j}
\label{cond1}
\end{equation}
where the transition rates $R_{ij}$ from state $j$ to state $i$ are time-independent, and $\sum_{j\neq i}R_{ji} = -R_{ii}$ is the escape rate from state $i$. The formal solution of (\ref{cond1}) reads
\begin{equation}
\mathbf{p}(\tau) = \mathrm{e}^{(\tau - \tau_0) \, R}\, \mathbf{p}(\tau_0) \,  ,
\label{cond2}
\end{equation}
where $\mathbf{p} \equiv (p_{1},\cdots,p_{K})^{T}$ and $\mathbf{p}(\tau_0)$ is an initial probability distribution.

We assume that the $K$ states form a connected network: from any state $i$ the system can evolve to any other state $j$ by a finite sequence of transitions.
We also assume that if $R_{ij} \neq 0$, then $R_{ji} \neq 0$.
These assumptions imply a unique stationary distribution, which we will denote by $\boldsymbol\pi = (\pi_1,\cdots,\pi_K)^T$:
\begin{equation}
\lim_{\tau\rightarrow\infty} \mathbf{p}(\tau) = \boldsymbol\pi \, .
\end{equation}
This distribution is characterized by stationary currents,
\begin{equation}
J_{ij} = R_{ij}\pi_j - R_{ji}\pi_i
\end{equation}
representing the flow of probability from $j$ to $i$ in the stationary state.
 
We denote by $\mathcal{P}[\mathcal{S}]$ the probability of observing a trajectory $\mathcal{S}$ over an interval of total duration $\mathcal{T}$, beginning at time $\tau_0 = -\mathcal{T}/2$ and ending at time $\mathcal{T}/2$.
Discretizing time in intervals of duration $\delta\tau$, we represent this trajectory as a sequence of states $\mathcal{S}=(s_{-N},s_{-N+1},\ldots, s_{N-1},s_{N}) \in \Lambda^{2N+1}$, where $s_{n}$ is the state of the system at time $n\,\delta\tau$ and $\delta\tau = \mathcal{T}/2N$.
We will take $\delta\tau$ to be infinitesimal, hence $\delta\tau\, R_{ij}\ll 1$ for all $i$, $j$, and we will ignore terms of order $\delta\tau^2$.
The probability $\mathcal{P}[\mathcal{S}]$ can be written as
\begin{equation}
\mathcal{P}[\mathcal{S}] 
= \left( \prod_{k=-N}^{N-1} U_{s_{k+1} s_{k}} \right) \, p_{s_{-N}}(\tau_0),
\label{cond3}
\end{equation}
where the transition probability $U_{i j}$ is given by
\begin{equation}
U_{i j} = \left(\mathrm{e}^{\delta\tau\, R} \right)_{i j} \simeq \delta_{i j} + \delta\tau\, R_{i j} \, .
\label{cond4}
\end{equation} 
The distribution $\mathcal{P}[\mathcal{S}]$ defines a statistical ensemble of discretized trajectories of duration $\mathcal{T}$.

The time-averaged entropy production rate along a trajectory $\mathcal{S}$ is given by
\begin{equation}
\sigma[\mathcal{S}] = \frac{1}{\mathcal{T}}\sum_{k=-N}^{N-1} \ln\left( \frac{R_{s_{k+1} s_{k}}}{R_{s_{k} s_{k+1}}} \right),
\label{cond5}
\end{equation}
where $\ln(R_{i j}/R_{j i})$ is the entropy production associated with a transition from $j$ to $i$ \cite{lebowitz1999,gaspard2004,seifert2005,imparato2007}.
The probability distribution of the time-averaged entropy production rate is then
\begin{equation}
P(\sigma) = \sum_{[\mathcal{S}]}\,\mathcal{P}[\mathcal{S}] \, \delta(\sigma - \sigma[\mathcal{S}]) \, ,
\label{cond6}
\end{equation}
where $\sum_{[\mathcal{S}]} \equiv \sum_{s_{-N},\cdots,s_{N}}$.
Finally, we define a {\it conditioned} ensemble of trajectories, described by a probability distribution $\mathcal{P}_c[\mathcal{S} | \sigma]$, which is the subensemble of $\mathcal{P}[\mathcal{S}]$ containing only those trajectories with a time-averaged entropy production rate equal to $\sigma$.  Our aim is to study the dynamics of the system within this conditioned ensemble.


Note that the distributions $\mathcal{P}[\mathcal{S}]$, $P(\sigma)$ and $\mathcal{P}_c[\mathcal{S} | \sigma]$ -- as well as $\mathcal{P}_\lambda[\mathcal{S}]$ and $P_\lambda(\sigma)$, defined below -- all depend on the duration $\mathcal{T}$, but this dependence is notationally suppressed.
We are interested in the long-time limit, $\mathcal{T}\rightarrow\infty$ (with $\delta\tau$ fixed, hence $N\rightarrow\infty$).
In this limit, the distribution $P(\sigma)$ becomes ever more sharply peaked around a value $\bar{\sigma}$, which is the infinite-time average entropy production rate in the stationary state:
\begin{equation}
\bar\sigma = \sum_{i>j} J_{ij} \ln \left( \frac{R_{ij}}{R_{ji}} \right) \, .
\label{eq:entropy-currentUNB}
\end{equation}

Now consider a {\it biased} ensemble, described by the probability distribution
\begin{equation}
\label{eq:lambdaEnsemble}
\mathcal{P}_\lambda[\mathcal{S}] = \frac{1}{\mathcal{N}} \mathcal{P}[\mathcal{S}] \, \mathrm{e}^{-\lambda \mathcal{T} \sigma[\mathcal{S}]}
\end{equation}
where $\lambda$ is a real parameter and
\begin{equation}
\label{eq:Ndef}
\mathcal{N} = \mathcal{N}(\lambda) = \sum_{[\mathcal{S}]}\,\mathcal{P}[\mathcal{S}] \mathrm{e}^{-\lambda \mathcal{T} \sigma[\mathcal{S}]}
\end{equation}
is a normalization factor.
The distribution $\mathcal{P}_\lambda[\mathcal{S}]$ defines a modified distribution of entropy production rates
\begin{equation}
P_{\lambda}(\sigma) = \sum_{[\mathcal{S}]}\,\mathcal{P}_{\lambda}[\mathcal{S}]\,\delta(\sigma - \sigma[\mathcal{S}]) .
\label{cond7}
\end{equation}
When $\lambda>0$, the factor $\mathrm{e}^{-\lambda \mathcal{T} \sigma[\mathcal{S}]}$ favors trajectories with low values of $\sigma$, hence the distribution $P_{\lambda}(\sigma)$ is shifted to the left of $P(\sigma)$; the opposite comments apply when $\lambda<0$.

As shown in Refs.~\cite{jack2010,chetrite2013,chetrite2014}, in the long-time limit the statistics of the biased ensemble become equivalent to those of a stationary Markov process, described by a rate matrix $\tilde R(\lambda)$, Eq.~(\ref{dual6}).
Moreover, in this limit the distribution $P_{\lambda}(\sigma)$ becomes sharply peaked around a value $\bar{\sigma}_\lambda$.
This value decreases monotonically with $\lambda$, and can be written as
\begin{equation}
\label{eq:barsigmalambda}
\bar\sigma_\lambda = \sum_{i>j} \tilde{J}_{ij}(\lambda) \ln \left( \frac{R_{ij}}{R_{ji}} \right) \, ,
\end{equation}
where $\tilde{J}_{ij}(\lambda)$ is the net flow of probability from $j$ to $i$ in the biased $\lambda$-ensemble, in the long-time limit.
Note that in Eqs.~(\ref{cond7}) and (\ref{eq:barsigmalambda}), the entropy production rate is defined with respect to the original, unbiased transition rates $R_{ij}$.

We can view the transformation $\mathcal{P}[\mathcal{S}]\to \mathcal{P}[\mathcal{S}] \mathrm{e}^{-\lambda \mathcal{T} \sigma[\mathcal{S}]}$ as a method for constructing ensembles of trajectories that are effectively conditioned on particular entropy production rates.
By varying $\lambda$, we ``tune in'' to trajectories with values of $\sigma$ near a desired value $\bar{\sigma}_\lambda$; the larger the value of $\mathcal{T}$, the narrower the distribution of values of $\sigma$ around $\bar{\sigma}_\lambda$.
Under conditions discussed by Jack and Sollich~\cite{jack2010} and by Chetrite and Touchette~\cite{chetrite2013,chetrite2014}, which are fulfilled by the model we study, in the long-time limit the ensemble $\mathcal{P}_\lambda[\mathcal{S}]$ becomes equivalent to an ensemble in which the entropy production rate is constrained to the value $\bar{\sigma}_\lambda$.
We represent this equivalence using the notation
\begin{equation}
\label{eq:equivalenceOfEnsembles}
\mathcal{P}_\lambda[\mathcal{S}] \sim \mathcal{P}_c[\mathcal{S} | \bar\sigma_\lambda] \, ,
\end{equation}
where the limit $\mathcal{T}\rightarrow\infty$ is implied.

As in the equilibrium case, the equivalence of nonequilibrium ensembles expressed by Eq.~(\ref{eq:equivalenceOfEnsembles}) helps us to avoid the difficulties imposed by sharp constraints. 
For nonequilibrium path-space ensembles, the long-time limit $\mathcal{T}\rightarrow\infty$ is analogous to the thermodynamic limit, and we can select a value of $\lambda$ that gives a desired entropy production rate $\bar{\sigma}_\lambda$:
the long-time dynamics of the system in the $\lambda$-ensemble $\mathcal{P}_\lambda[\mathcal{S}]$ are equivalent to its dynamics in the corresponding fixed-$\sigma$ ensemble, $\mathcal{P}_c[\mathcal{S} | \bar\sigma_\lambda]$.
We will exploit this equivalence in order to explore the behavior of the system when its long-time-averaged entropy production rate is conditioned on a particular value.

In the remainder of this section we first establish that the $\lambda$-ensemble $\mathcal{P}_\lambda[\mathcal{S}]$ describes a stationary process whose dynamics are Markovian in the long-time limit, Eq.~(\ref{eq:MarkovCondition}), and we obtain the rate matrix that generates this process, Eq.~(\ref{dual6}).

We begin with the expression
\begin{equation}
\mathcal{P}_{\lambda}[\mathcal{S}] = \mathcal{N}^{-1} \left( \prod_{k=-N}^{N-1} U_{s_{k+1} s_{k}} \right) \mathrm{e}^{-\lambda \mathcal{T} \sigma[S]}  p_{s_{-N}}(\tau_0) = \mathcal{N}^{-1} \left[ \prod_{k=-N}^{N-1} U_{s_{k+1} s_{k}} \left( \frac{R_{s_{k} s_{k+1}}}{R_{s_{k+1} s_{k}}} \right)^{\lambda} \right] p_{s_{-N}}(\tau_0),
\label{cond8}
\end{equation}
which follows from Eqs.~(\ref{cond3}) and (\ref{eq:lambdaEnsemble}), along with
\begin{equation}
\mathrm{e}^{-\lambda \mathcal{T} \sigma[\mathcal{S}]} = \prod_{k=-N}^{N-1} \left( \frac{R_{s_{k} s_{k+1}}}{R_{s_{k+1} s_{k}}} \right)^{\lambda}.
\label{cond9}
\end{equation}
Defining a matrix $Q(\lambda)$ with elements $Q_{i j} \equiv U_{i j} (R_{j i}/R_{i j})^{\lambda}$, we rewrite Eq.~(\ref{cond8}) as
\begin{equation}
\label{cond9.5}
\mathcal{P}_{\lambda}[\mathcal{S}] = \mathcal{N}^{-1} \left[ \prod_{k=-N}^{N-1} Q_{s_{k+1} s_{k}}  \right] p_{s_{-N}}(\tau_0).
\end{equation}
Moreover we can use Eq.~(\ref{cond4}) to obtain
\begin{equation}
Q_{i j}\simeq  \left\{
\begin{aligned}
1 + R_{i i}\, \delta t,&\;\;i=j \\
R_{i j}^{1-\lambda} R_{j i}^{\lambda}\, \delta t,&\;\; i\neq j
\end{aligned}
\right.\,.
\label{cond10}
\end{equation}

Next, we introduce the convenient bra-ket notation to denote the right and left eigenvectors of $Q$:
\begin{equation}
Q | \mathbf{u}_{k} \rangle = \alpha_{k} | \mathbf{u}_{k} \rangle,
\qquad,\qquad
\langle \mathbf{v}_{k}| Q = \alpha_{k} \langle \mathbf{v}_{k} |,
\label{cond13}
\end{equation}
where $\alpha_{k}$ are the corresponding eigenvalues. For the sake of simplicity we assume $Q$ to be diagonalizable, so that the following relations apply after suitable normalization of the eigenvectors:
\begin{subequations}
\begin{align}
\langle \mathbf{v}_{k} | \mathbf{u}_{j} \rangle = \delta_{k j}\,, \\
\sum_{k} |\mathbf{u}_{k} \rangle\langle \mathbf{v}_{k}| = \mathds{1}\,.
\end{align}
\label{cond13b}
\end{subequations}
In the appendix we extend our derivation to the case when $Q$ is {\it defective}, that is non-diagonalizable \cite{shilov}.

Because the matrix $Q$ is non-negative (by construction) and irreducible (since the $K$ states form a connected network), the conditions of the Perron-Frobenius theorem are satisfied \cite{seneta}.
This theorem tells us that $Q$ has a real eigenvalue $\mu$ whose value is greater than the modulus of any other eigenvalue:
\begin{equation}
\label{eq:alphas}
\mu \equiv \alpha_1 > \vert \alpha_{2}\vert  \ge \vert \alpha_{3}\vert \ge \ldots.
\end{equation}
Without loss of generality we have arranged the $\alpha_k$'s in descending order of their moduli.
We denote by
\begin{equation}
\label{eq:psiomega}
| \boldsymbol\psi\rangle \equiv |\mathbf{u}_{1}\rangle
\qquad {\rm and} \qquad
\langle\boldsymbol\omega| \equiv \langle \mathbf{v}_{1}|
\end{equation}
the right and left eigenvectors associated with the eigenvalue $\mu$.
The Perron-Frobenius theorem further guarantees that the elements of these eigenvectors are strictly positive.
Note that the eigenvectors and eigenvalues of $Q(\lambda)$ depend on the value of $\lambda$.
In particular, when $\lambda=0$ we have $Q=U$, $\mu=1$, $|\boldsymbol\psi\rangle = | \boldsymbol\pi\rangle$ and $\langle \boldsymbol\omega | = \langle \mathbf{1}|$, where $\langle \mathbf{1}|\equiv (1, 1,\ldots, 1)$.

From Eq.~(\ref{cond9.5}) we have
\begin{eqnarray}
\mathcal{N} &=&  \sum_{[\mathcal{S}]}\,\left[ \prod_{k=-N}^{N-1} Q_{s_{k+1} s_{k}}  \right] p_{s_{-N}}(\tau_0) \\
&=& \sum_{[\mathcal{S}]}\, Q_{s_N s_{N-1}} Q_{s_{N-1} s_{N-2}} \cdots Q_{s_{-N+1} s_{-N}} p_{s_{-N}}(\tau_0) \\
&=& 
\langle \mathbf{1} \vert Q^{2N} \vert \mathbf{p}(\tau_0)\rangle
\end{eqnarray}
Introducing the notation $\mathbf{p}_0 \equiv \mathbf{p}(\tau_0)$ and using Eq.~(\ref{cond13b}), we obtain
\begin{equation}
\mathcal{N} = \sum_k \langle \mathbf{1} \vert Q^{2N} \vert \mathbf{u}_{k} \rangle\langle \mathbf{v}_{k} \vert \mathbf{p}_0\rangle
= \sum_k \alpha_k^{2N} \langle \mathbf{1} \vert \mathbf{u}_{k} \rangle\langle \mathbf{v}_{k} \vert \mathbf{p}_0\rangle
\rightarrow \mu^{2N} \langle \mathbf{1} \vert \boldsymbol\psi \rangle\langle \boldsymbol\omega \vert \mathbf{p}_0\rangle \quad ,
\end{equation}
where $\rightarrow$ denotes the long-time limit ($\mathcal{T},N \to \infty$), and we have invoked Eq.~(\ref{eq:alphas}) when taking this limit.

With these elements in place, let us consider the probability, in the $\lambda$-ensemble, that the system proceeds through a sequence of states $r_0, r_1, \cdots r_\kappa$ at times $0, \delta\tau, \cdots \kappa\delta\tau$.
(Without loss of generality, we have taken $j=0$ as the initial time step.)
This probability is given by
\begin{equation}
p_\lambda(r_{\kappa},\kappa; r_{\kappa - 1}, \kappa - 1;\ldots ;r_{0},0) =
\mathcal{N}^{-1}
\sum_{[\mathcal{S}]}\, \left[ \prod_{k=-N}^{N-1} Q_{s_{k+1} s_{k}}  \right] p_{s_{-N}}(\tau_0) \, \delta_{s_0 r_0}, \delta_{s_1 r_1} \cdots \delta_{s_\kappa r_\kappa} .
\end{equation}
Here we have inserted appropriate Kronecker delta functions $\delta_{s_j r_j}$ into Eq.~(\ref{cond9.5}) and taken a sum over trajectories.
Let us now introduce the notation $|\mathbf{e}_{s}\rangle = (\delta_{1 s},\delta_{2 s},\ldots,\delta_{K s})^{T}$ to denote a unit vector.
Proceeding as in the previous paragraph we get
\begin{eqnarray}
p_\lambda(r_{\kappa},\kappa; \ldots ;r_{0},0) &=&
\mathcal{N}^{-1} \langle \mathbf{1} \vert Q^{N-\kappa} \vert \mathbf{e}_{r_\kappa} \rangle Q_{r_\kappa r_{\kappa-1}} \cdots Q_{r_1 r_0} \langle \mathbf{e}_{r_0} \vert Q^N \vert \mathbf{p}_0 \rangle \nonumber \\
 &=& \mathcal{N}^{-1} \sum_{l,m} \langle \mathbf{1} \vert Q^{N-\kappa} \vert \mathbf{u}_{l} \rangle\langle \mathbf{v}_{l} \vert \mathbf{e}_{r_\kappa} \rangle
 Q_{r_\kappa r_{\kappa-1}} \cdots Q_{r_1 r_0} \langle \mathbf{e}_{r_0} \vert Q^N \vert \mathbf{u}_{m} \rangle\langle \mathbf{v}_{m} \vert \mathbf{p}_0 \rangle \nonumber \\
 &\rightarrow& \mathcal{N}^{-1} \mu^{2N-\kappa} \langle \mathbf{1} \vert \boldsymbol\psi \rangle \omega_{r_\kappa}
 \left(\prod_{j=0}^{\kappa-1} Q_{r_{j+1} r_{j}} \right) \psi_{r_0} \langle \boldsymbol\omega \vert \mathbf{p}_0\rangle \nonumber \\
 &=& \mu^{-\kappa} \omega_{r_\kappa} \left(\prod_{j=0}^{\kappa-1} Q_{r_{j+1} r_{j}} \right) \psi_{r_0}
 = \left( \prod_{j=0}^{\kappa - 1}\frac{\omega_{r_{j+1}}\,Q_{r_{j+1} r_{j}}\,\omega^{-1}_{r_{j}}}{\mu} \right)\,\omega_{r_{0}} \psi_{r_{0}}
 \label{eq:jointPD}
\end{eqnarray}
For the special cases $\kappa = 0$ and $\kappa = 1$ we get
\begin{equation}
\label{eq:specialCases}
p_\lambda(r_{0},0) = \omega_{r_{0}} \psi_{r_{0}}
\qquad,\qquad
p_\lambda(r_1, 1; r_0,0) = \frac{\omega_{r_1}\,Q_{r_1 r_0}\,\omega^{-1}_{r_0}}{\mu} \,\omega_{r_{0}} \psi_{r_{0}} \quad ,
\end{equation}
implying that the conditional probability to be found in state $r_1$ at $t=\delta\tau$, given state $r_0$ at $t=0$, is
\begin{equation}
\label{eq:condPD}
p_\lambda(r_1, 1 \vert r_0,0) = \frac{\omega_{r_1}\,Q_{r_1 r_0}\,\omega^{-1}_{r_0}}{\mu} .
\end{equation}
An identical result holds for $p_\lambda(r_{j+1},j+1 \vert r_j,j)$ for any $j$ (since the choice of $j=0$ as the initial time step in Eq.~(\ref{eq:jointPD}) was arbitrary), allowing us to rewrite Eq.~(\ref{eq:jointPD}) as follows:
\begin{equation}
\label{eq:MarkovCondition}
p_\lambda(r_{\kappa},\kappa; \ldots ;r_{0},0) = \left[ \prod_{j=0}^{\kappa - 1} p_{\lambda}(r_{j+1}, j+1 | r_{j}, j) \right] p_{\lambda}(r_{0},0) ,
\end{equation}
from which we conclude that the trajectory segment $r_0, r_1, \cdots r_\kappa$, sampled from the $\lambda$-ensemble, is described statistically by a Markov process, in the limit $\mathcal{T},N \to \infty$.

Following Refs.~\cite{jack2010,chetrite2013,chetrite2014}, let us now construct the rate matrix that governs the $\lambda$-biased dynamics.
As discussed in Refs.~\cite{chetrite2013,chetrite2014}, this rate matrix is related to the original, unbiased rate matrix by Doob's transform \cite{doob1984}.

We have used the notation $R$ and $U$ to denote the rate matrix and the single-time-step transition matrix for the unbiased dynamics.
In what follows we will use $\tilde R(\lambda)$ and $\tilde U(\lambda)$ to denote the rate and transition matrices for the biased dynamics, thus $\tilde R(0)=R$ and $\tilde U(0)=U$.
The elements of $\tilde U(\lambda)$ are given by Eq.~(\ref{eq:condPD}):
\begin{equation}
\label{eq:tildeT}
\tilde U = \frac{1}{\mu} \Omega Q \Omega^{-1}
\end{equation}
where $\Omega \equiv \mathrm{diag}(\langle\boldsymbol\omega |)$ is a diagonal matrix whose elements are the components of the left eigenvector 
$\langle\boldsymbol\omega |$ of $Q$.
Using $\tilde U = e^{\tilde R\delta\tau} \simeq \mathds{1} + \tilde R\delta\tau$ (where $\mathds{1}$ is the identity matrix) we obtain
\begin{equation}
\label{dual1}
\tilde R \delta\tau = \frac{1}{\mu} \Omega Q \Omega^{-1} - \mathds{1}
\end{equation}
To rewrite this expression in a more convenient form, we first use Eq.~(\ref{cond10}) to write
\begin{equation}
Q = \mathds{1} + M\,\delta\tau \,,
\label{dual2}
\end{equation}
where
\begin{equation}
M_{i j} = \left\{
\begin{aligned}
 R_{i i}\,,&\;\; i=j \\
 R_{i j}^{1-\lambda} R_{j i}^{\lambda}\,,&\;\; i\neq j \,.
\end{aligned}
\right.
\label{dual3}
\end{equation}
By Eq.~(\ref{dual2}), $Q$ and $M$ have the same eigenvectors but different eigenvalues, and in particular
\begin{equation}
\mu = 1 + \eta\,\delta\tau \quad,\quad \langle \boldsymbol\omega| M = \eta \langle \boldsymbol\omega| \quad,\quad M | \boldsymbol\psi\rangle = \eta | \boldsymbol\psi\rangle \, .
\label{dual4}
\end{equation}
Keeping terms only up to first order in $\delta\tau$, Eq.~(\ref{dual1}) gives us
\begin{equation}
\tilde{R}(\lambda) = \Omega\, M\,\Omega^{-1} - \eta \mathds{1} \, ,
\label{dual6}
\end{equation}
where the quantities appearing on the right side of the equation depend on $\lambda$, but not on $\delta\tau$.
This rate matrix, $\tilde{R}(\lambda)$, generates trajectories with the same statistics as those of the $\lambda$-ensemble, in the long-time limit.

Let us now solve for the null right eigenvector of the rate matrix $\tilde{R}(\lambda)$.
We will use the notation $| \tilde{\boldsymbol\pi}(\lambda) \rangle$ to denote this eigenvector,
whose components $\tilde{\pi}_s = \langle \mathbf{e}_s | \tilde{\boldsymbol\pi} \rangle$ represent the stationary probability distribution in the biased $\lambda$-ensemble.
We have 
\begin{equation}
| \mathbf{0} \rangle = 
\tilde{R} | \tilde{\boldsymbol\pi} \rangle
= \Omega M \Omega^{-1} | \tilde{\boldsymbol\pi} \rangle - \eta | \tilde{\boldsymbol\pi} \rangle
\end{equation}
hence
\begin{equation}
M \, \Bigl( \Omega^{-1} | \tilde{\boldsymbol\pi} \rangle \Bigr)  = \eta \, \Bigl(  \Omega^{-1} | \tilde{\boldsymbol\pi} \rangle \Bigr) \, ,
\end{equation}
which implies by Eq.~(\ref{dual4}) that $\Omega^{-1} | \tilde{\boldsymbol\pi} \rangle \propto | \boldsymbol\psi\rangle$.
We thus have $| \tilde{\boldsymbol\pi}\rangle = c\,\Omega | \boldsymbol\psi\rangle$, or $\tilde{\pi}_s = \langle \mathbf{e}_{s} | \tilde{\boldsymbol\pi} \rangle = c \, \omega_s \psi_s$.
The constant $c$ is set by the normalization condition $\sum_s \tilde\pi_s = 1$, which combines with $\langle \mathbf{v}_{1}| \mathbf{u}_{1}\rangle = 1$ to give us $c=1$, hence
\begin{equation}
\label{eq:pitildepi}
| \tilde{\boldsymbol\pi}\rangle = \Omega | \boldsymbol\psi\rangle
\qquad,\qquad
\tilde{\pi}_s = \omega_s \psi_s \, .
\end{equation}
This is consistent with Eq.~(\ref{eq:specialCases}).

To end this section, we establish symmetry identities, Eqs.~(\ref{eq:eigensymmetry}) and (\ref{eq:RSsymmetry}), that will allows us to investigate the dependence of the biased dynamics on the parameter $\lambda$.
From Eq.~(\ref{dual3}) we have
$M(1-\lambda) = M(\lambda)^T$,
which implies 
\begin{equation}
\label{eq:eigensymmetry}
\eta(\lambda) = \eta(1-\lambda)
\quad,\quad
| \boldsymbol\psi(\lambda) \rangle = | \boldsymbol\omega(1-\lambda) \rangle \, .
\end{equation}
These results combine with the transpose of Eq.~(\ref{dual6}) to give us
\begin{equation}
\label{eq:RSsymmetry}
\Omega(\lambda) \tilde{R}(\lambda)^T \Omega(\lambda)^{-1} = M(1-\lambda) - \eta(1-\lambda) \mathds{1}
\end{equation}
The symmetry property embodied by Eqs.~(\ref{eq:eigensymmetry}) and (\ref{eq:RSsymmetry}) was previously obtained by Lebowitz and Spohn~\cite{lebowitz1999}, in their proof of the fluctuation theorem for general Markov processes.
A similar symmetry is implicit in Kurchan's earlier derivation of the fluctuation theorem for Langevin processes~\cite{kurchan1998}.

\section{Dual dynamics and conditional Reversibility \label{sec:rever}}

We are now in a position to study the {\it dual} of the rate matrix $\tilde R$, defined by \cite{kemeny,crooks2000}
\begin{equation}
\label{eq:dualDef}
\tilde{R}^{\rm dual} = \tilde{\Pi} \tilde{R}^T \tilde{\Pi}^{-1}
\end{equation}
where $\tilde{\Pi} = \rm{diag} ( | \tilde{\boldsymbol\pi}\rangle )$.
From this definition it follows that $\langle\mathbf{1} | \tilde{R}^{\rm dual} = \langle\mathbf{0}|$ and $\tilde{R}^{\rm dual} | \tilde{\boldsymbol\pi} \rangle = | \mathbf{0}\rangle$, hence $\tilde{R}^{\rm dual}$ is a rate matrix whose stationary distribution is the same as that of $\tilde{R}$.
Moreover, in the stationary state the flow of probability under the dual dynamics is exactly opposite to the probability flow under the original dynamics:
\begin{equation}
\label{eq:dualCurrents}
\tilde{J}_{ij}^{\rm dual} = \tilde{R}_{ij}^{\rm dual} \tilde{\pi}_j - \tilde{R}_{ji}^{\rm dual} \tilde{\pi}_i 
= \left( \tilde{\pi}_i \tilde{R}_{ji} \tilde{\pi}_j^{-1} \right) \tilde{\pi}_j - \left( \tilde{\pi}_j \tilde{R}_{ij} \tilde{\pi}_i^{-1} \right) \tilde{\pi}_i = -\tilde{J}_{ij} \, .
\end{equation}

Combining Eq.~(\ref{eq:dualDef}) with Eqs.~(\ref{dual6}), (\ref{eq:pitildepi}), (\ref{eq:eigensymmetry}) and (\ref{eq:RSsymmetry}), we obtain
\begin{eqnarray}
\left[ \tilde{R}^{\rm dual}(\lambda) \right]_{ij} &=& \tilde{\pi}_{i}(\lambda) \left[ \tilde{R}(\lambda)^T \right]_{ij} \tilde{\pi}_j(\lambda)^{-1} \nonumber \\
&=& \psi_i(\lambda) \left[ \Omega(\lambda) \tilde{R}(\lambda)^T \Omega(\lambda)^{-1} \right]_{ij} \psi_j(\lambda)^{-1} \nonumber \\
&=& \psi_i(\lambda) \left[ M(1-\lambda) - \eta(1-\lambda) \mathds{1} \right]_{ij} \psi_j(\lambda)^{-1} \nonumber \\
&=& \omega_i(1-\lambda) \left[ M(1-\lambda) \right]_{ij} \omega_j(1-\lambda)^{-1} - \eta(1-\lambda) \delta_{ij} = \left[ \tilde{R}(1-\lambda) \right]_{ij}
\end{eqnarray}
That is,
\begin{equation}
\label{eq:dualMainResult}
\tilde{R}^{\rm dual}(\lambda) = \tilde{R}(1-\lambda) \, .
\end{equation}
Together with Eq.~(\ref{eq:dualCurrents}), this result gives us
\begin{equation}
\tilde{J}_{ij}(1-\lambda) = - \tilde{J}_{ij}(\lambda) \, .
\label{eq:currentsTR}
\end{equation}
Thus the stationary currents in the $\lambda$-ensemble are the opposite of those in the $(1-\lambda)$-ensemble, which further implies (see Eq.~(\ref{eq:barsigmalambda})) that the corresponding average entropy production rates are opposite:
\begin{equation}
\bar\sigma_{1-\lambda} = -\bar\sigma_\lambda \, .
\label{eq:entropyTR}
\end{equation}
Therefore, by the equivalence of ensembles expressed by Eq.~(\ref{eq:equivalenceOfEnsembles}), in the long-time limit the biased ensembles $\mathcal{P}_{\lambda}[\mathcal{S}]$ and $\mathcal{P}_{1-\lambda}[\mathcal{S}]$ become equivalent to ensembles {\it conditioned on opposite entropy production rates}, $\mathcal{P}_c[\mathcal{S} | \bar\sigma_\lambda]$ and
$\mathcal{P}_c[\mathcal{S} | \!-\!\bar\sigma_\lambda]$.

In Sec.~\ref{sec:cond} we discussed the equivalence between {\it biased} and {\it conditioned} ensembles of trajectories; see Eq.~(\ref{eq:equivalenceOfEnsembles}). 
Combined with Eq.~(\ref{eq:dualMainResult}), this equivalence implies that ensembles conditioned on opposite values of entropy production, $\mathcal{P}_c[\mathcal{S} | \sigma]$ and $\mathcal{P}_c[\mathcal{S} | \!-\!\sigma]$, are characterized by the same stationary distribution, but opposite currents.
This suggests that these ensembles may be related by time-reversal, not only with respect to time-averaged currents and entropy production, but also at the level of individual trajectories.
We now explore this idea in detail, to arrive at our central result, Eq.~(\ref{eq:condRev}).

Let us consider an interval of time of finite duration $\kappa\,\delta\tau$, and 
let ${\bf r} = (r_0 \rightarrow r_1 \rightarrow \cdots \rightarrow r_\kappa)$ denote
a trajectory segment evolving during this interval, in which the system is found in state $r_0$ at time step $0$, in state $r_1$ at time step $1$, and so forth up to time step $\kappa$.
The notation ${\bf r}^* = (r_\kappa \rightarrow \cdots \rightarrow r_1 \rightarrow r_0)$ will denote the time-reversed segment, in which the same states are visited in reverse order.
If we sample a trajectory $\mathcal{S}$ from the biased $\lambda$-ensemble, and we examine the states visited by this trajectory from $\tau=0$ to $\tau=\kappa\,\delta\tau$, then the probability to observe the trajectory segment ${\bf r}$ is given by
\begin{eqnarray}
\label{eq:Pforward}
P_\lambda({\bf r})
&=& \tilde{\pi}_{r_0}(\lambda) \tilde{U}_{r_1r_0}(\lambda) \cdots \tilde{U}_{r_\kappa r_{\kappa-1}}(\lambda) \nonumber\\
&\rightarrow& \left[ \frac{1}{\mu(\lambda)} \right]^\kappa \, \omega_{r_\kappa}(\lambda) \left[ \prod_{j=0}^{\kappa-1} Q_{r_{j+1} r_j}(\lambda) \right] \psi_{r_0}(\lambda) \, ,
\end{eqnarray}
using Eqs.~(\ref{eq:tildeT}) and (\ref{eq:pitildepi}).
Similarly,
\begin{equation}
\label{eq:Preverse}
P_{1-\lambda}({\bf r}^*)
\rightarrow \left[ \frac{1}{\mu(1-\lambda)} \right]^\kappa \, \omega_{r_0}(1-\lambda) \left[ \prod_{j=0}^{\kappa-1} Q_{r_{j} r_{j+1}}(1-\lambda) \right] \psi_{r_\kappa}(1-\lambda) \, .
\end{equation}
Let us now compare these probabilities.

From Eq.~(\ref{cond10}) we have
$Q(1-\lambda)  = Q(\lambda)^T$, which implies \cite{lebowitz1999}
\begin{equation}
\mu(1-\lambda) = \mu(\lambda)
\quad,\quad
\omega_s(1-\lambda) = \psi_s(\lambda)
\end{equation}
(compare with Eq.~(\ref{eq:eigensymmetry})).
Combining these results with Eqs.~(\ref{eq:Pforward}) and (\ref{eq:Preverse}), we obtain
\begin{equation}
P_\lambda({\bf r}) = P_{1-\lambda}({\bf r}^*) \, .
\label{eq:biasedETR}
\end{equation}
By the equivalence of ensembles, this implies the following conclusion.
The probability to observe a given trajectory segment ${\bf r} = (r_0 \rightarrow \cdots \rightarrow r_\kappa)$ when conditioning on a particular value of time-averaged entropy production rate $\sigma$, is the same as the probability to observe the time-reversed segment ${\bf r}^* = (r_\kappa \rightarrow \cdots \rightarrow r_0)$ when conditioning on the opposite value of time-averaged entropy production rate, $-\sigma$.
Using obvious notation:
\begin{equation}
\label{eq:condRev}
P_c({\bf r} | \sigma) = P_c({\bf r}^* |\!-\!\sigma)
\end{equation}
It is in this sense that the two conditioned ensembles are related by time-reversal.

Equation (\ref{eq:biasedETR}) is related to expressions that already exist in the literature. For instance, Eq.~(37) of Ref.~\cite{gaspard2004} and Eq.~(2.21) of Ref.~\cite{lebowitz1999} can be written in our notation as (see Sec.~\ref{sec:cond})
\begin{equation}
\label{eq:existing}
\mathcal{P}[\mathcal{S}]\, e^{-\lambda \mathcal{T}\sigma[\mathcal{S}]} = \frac{p_{s_{-N}}(\tau_{0})}{p_{s_{N}}(\tau_{f})} \mathcal{P}[\mathcal{S}^{*}]\, e^{-(1-\lambda) \mathcal{T}\sigma[\mathcal{S}^{*}]}\,,
\end{equation}
where $\tau_{f} - \tau_{0} = \mathcal{T}$ is the total duration of the trajectory $\mathcal{S}$, $s_{-N}$ and $s_{N}$ are the corresponding initial and final states and $\mathcal{S}^{*}$ is the time reversal of $\mathcal{S}$. We recognize in this expression the {\it unnormalized} distributions of the $\lambda$ and $(1-\lambda)$-ensembles (see Eq.~(\ref{eq:lambdaEnsemble})). In the long-time limit $\mathcal{T}\rightarrow\infty$, the first factor appearing on the right side of Eq.~(\ref{eq:existing}) can be neglected (it is subdominant), leading to Eq.~(\ref{eq:biasedETR}). For stochastic systems governed by time-dependent rate matrices, a transient result analogous to our Eq.~(\ref{eq:biasedETR}) appears as Eq.~(3.35) in Ref.~\cite{harris2007}.
However, to the best of our knowledge, the symmetry expressed by Eq. (\ref{eq:biasedETR}) has not previously been combined with the equivalence of ensembles (Eq. (\ref{eq:equivalenceOfEnsembles})) to arrive at our central result, Eq. (\ref{eq:condRev}), which has a simple and appealing physical interpretation.

Equation (\ref{eq:condRev}) is the stochastic version of Gallavotti's {\it conditional reversibility theorem} \cite{gallavotti1999}.
Gallavotti's theorem is derived in the context of deterministic dynamics and is stated in terms of ``fluctuation patterns'', a term that denotes the evolution of an arbitrary observable over a finite interval of time.
However the basic content of the theorem is the same as that of Eq.~(\ref{eq:condRev}): the probability to observe a system to behave in a particular manner when conditioning on one value of entropy production, is the same as the probability to observe the time-reversed behavior when conditioning on the opposite value.
In particular, suppose we condition on the value $-\bar\sigma$, which is the opposite of the entropy production rate in the unbiased ensemble. 
Then Eq.~(\ref{eq:condRev}) implies that the trajectories we will observe are statistically equivalent, under time-reversal, to the those generated by the original, unbiased dynamics, Eq.~(\ref{cond1}).
In effect, in the ensemble conditioned on $\sigma = -\bar\sigma$, time will appear to be running backward.
An analogous result has been obtained for the case of systems driven away from equilibrium by varying a parameter of the Hamiltonian~\cite{jarzynski2006,jarzynski2011}, rather than by equations of motion that violate detailed balance.

\section{Example \label{sec:example}}

\begin{figure}
\includegraphics[width=0.48\textwidth]{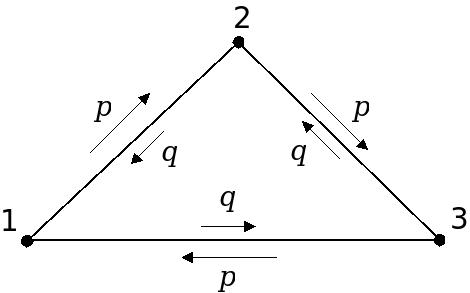}
\caption{Schematic representation of the transitions described by the matrix $R$.
Clockwise transitions occur at an average rate $p$, and counterclockwise transitions at an average rate $q$.
\label{fig1}}
\end{figure}

We now illustrate our results with an exactly solvable model. Consider the rate matrix
\begin{equation}
R = 
\begin{pmatrix}
-1 & q & p \\
p & -1 & q \\
q & p & -1 \\
\end{pmatrix}\,,
\label{eq:matrixR}
\end{equation}
whose off-diagonal elements represent the transitions rates of a system with only three states, as shown schematically in Fig.~\ref{fig1}. We impose $p + q = 1$ in order to fulfill the condition $\sum_i R_{i j}=0$. 
(Note that $q$ and $p$ are {\it rates}, thus the condition $p+q=1$ reflects our choice to set the escape rate from any state $i$ to unity: $| R_{ii} | = 1$.)

We start by calculating the stationary currents, $J_{i j}$, and the infinite-time average entropy production rate, $\bar{\sigma}$, in the unbiased ensemble. The stationary distribution in this case is $| \boldsymbol\pi \rangle = (1/3,1/3,1/3)^{T}$, from which we first obtain the currents,
\begin{equation}
J = \frac{1}{3}
\begin{pmatrix}
0 & (q-p) & (p-q) \\
(p-q) & 0 & (q-p) \\
(q-p) & (p-q) & 0\\
\end{pmatrix}\,,
\end{equation}
and then the average entropy production rate, using Eq.~(\ref{eq:entropy-currentUNB}):
\begin{equation}
\bar{\sigma} = (p-q) \ln{\left(\frac{p}{q}\right)}\, .
\end{equation}

Analogously, the stationary currents $\tilde{J}_{i j}(\lambda)$ and the infinite-time average entropy production rate $\bar{\sigma}_{\lambda}$ for a biased ensemble are obtained from the transition rate matrix $\tilde{R}(\lambda)$. From Eq.~(\ref{dual6}), this matrix is constructed from the matrix $M(\lambda)$ (see Eq.~(\ref{dual3})), 
\begin{equation}
M(\lambda) =
\begin{pmatrix}
-1 & q\left(\frac{p}{q} \right)^{\lambda} & p\left(\frac{q}{p} \right)^{\lambda} \\
p\left(\frac{q}{p} \right)^{\lambda} & -1 & q\left(\frac{p}{q} \right)^{\lambda} \\
q\left(\frac{p}{q} \right)^{\lambda}  & p\left(\frac{q}{p} \right)^{\lambda} & -1 \\ 
\end{pmatrix}
\equiv
\begin{pmatrix}
-1 & q_\lambda & p_\lambda\\
p_\lambda & -1 & q_\lambda \\
q_\lambda & p_\lambda & -1 \\
\end{pmatrix}\, ,
\label{eq:matrixS3st}
\end{equation}
its largest eigenvalue, 
\begin{equation}
\eta(\lambda) = -1 + q_\lambda + p_\lambda \,, 
\label{eq:eigvlS3st}
\end{equation}
and the corresponding left-eigenvector, 
\begin{equation}
\langle\boldsymbol\omega(\lambda)| = \left( 1, 1, 1 \right)\,.
\label{eq:leigvtS3st} 
\end{equation}
Using Eqs.~(\ref{eq:matrixR}), (\ref{eq:matrixS3st}), (\ref{eq:eigvlS3st}) and (\ref{eq:leigvtS3st}) in Eq.~(\ref{dual6}), we obtain
\begin{equation}
\tilde{R}(\lambda) = 
\begin{pmatrix}
-(q_\lambda+p_\lambda) & q_\lambda & p_\lambda\\
p_\lambda & -(q_\lambda+p_\lambda) & q_\lambda \\
q_\lambda & p_\lambda & -(q_\lambda+p_\lambda) \\
\end{pmatrix}\, .
\label{eq:R3st}
\end{equation}

The stationary distribution of (\ref{eq:R3st}) is $| \tilde{\boldsymbol\pi}(\lambda) \rangle = (1/3, 1/3, 1/3)^{T}$, from which we obtain the stationary currents
\begin{equation}
\tilde{J}(\lambda) = \frac{1}{3}
\begin{pmatrix}
0 & (q_\lambda - p_\lambda) & (p_\lambda - q_\lambda) \\
(p_\lambda - q_\lambda) & 0 & (q_\lambda - p_\lambda) \\
(q_\lambda - p_\lambda) & (p_\lambda - q_\lambda) & 0 \\
\end{pmatrix}\,,
\label{eq:currentlambda}
\end{equation}
and the average entropy production rate,
\begin{equation}
\bar{\sigma}_{\lambda} = \left( p_\lambda - q_\lambda \right) \ln{\left( \frac{p}{q}\right)}\,.
\label{eq:sigmalambda}
\end{equation}
Finally, note that Eq.~(\ref{eq:matrixS3st}) implies the identity
\begin{equation}
\label{eq:identity}
q_\lambda = p_{1-\lambda} \, .
\end{equation}

From Eq.~\ref{eq:R3st} we see that in the biased ensemble, the system makes clockwise transitions at an average rate $p_\lambda$ and counterclockwise transitions at a rate $q_\lambda$ (compare with Fig.~\ref{fig1}).
Combining this observation with Eq.~\ref{eq:identity}, we conclude that the dynamics in the $\lambda$-ensemble are the time-reversal of those in the $(1-\lambda)$-ensemble: the clockwise transition rates in one case become the counterclockwise rates in the other case.
In particular, for $\lambda=1$ the biased dynamics are obtained from the unbiased, $\lambda=0$ dynamics by exchanging the roles of $q$ and $p$ in the rate matrix $R$.
Moreover, Eqs.~(\ref{eq:currentsTR}) and (\ref{eq:entropyTR}) are easily verified for an arbitrary value of $\lambda$, using Eqs.~(\ref{eq:currentlambda}) - (\ref{eq:identity}).

\section{Conclusions \label{sec:conclu}}

The equivalence of ensembles in the thermodynamic limit is a familiar and important concept in equilibrium statistical physics.
In recent years analogous principles of equivalence have been developed for ensembles of trajectories representing systems away from thermal equilibrium \cite{evans2004,evans2005,maes2008,evans2010,jack2010,chetrite2013,chetrite2014}, with the thermodynamic limit replaced by the long-time limit.
In this paper we have applied these ideas to study discrete-state Markov processes that are conditioned on values of the time-averaged rate of entropy production.
We first mapped the original Markov process onto a new, biased Markov process that describes the conditioned ensemble, as in Refs.~\cite{jack2010,chetrite2013,chetrite2014}.
We then used the properties of the biased transition rate matrix to establish our central result, which states that trajectories conditioned on opposite entropy production rates are related statistically by time-reversal.
This extends Gallavotti's conditional reversibility theorem \cite{gallavotti1999}, originally formulated for deterministic dynamics, to the case of stochastic dynamics.

We end by pointing out that the study of conditioned ensembles has often led to insights in nonequilibrium statistical physics.
Most prominently, in two groundbreaking papers \cite{onsager1931a,onsager1931b} Onsager considered the spontaneous fluctuations of a system in equilibrium.
By focusing on rare fluctuations that produce an ``asymmetric distribution of energy'' -- or some other condition ordinarily associated with a system that is deliberately prepared away from equilibrium -- Onsager was led to the {\it regression hypothesis} and the {\it reciprocal relations}, which lie at the foundations of linear response theory and the fluctuation-dissipation theorem.
In recent years, Bertini {\it et al} \cite{bertini2001,bertini2002,bertini2015} have developed {\it macroscopic fluctuation theory}, which builds on a large deviation-like formula for space-time fluctuations and considerably extends Onsager's approach.
Separately, Rahav and Jarzynski \cite{rahav2013} have argued that when Onsager's arguments are extended beyond the regime of linear response, they lead naturally to far-from-equilibrium {\it fluctuation theorems}.
The common thread in these studies is a focus on spontaneous fluctuations conditioned on rare values of selected observables, such as energy distributions~\cite{onsager1931a} or currents density profiles~\cite{bertini2015}.
In the present paper, by considering rare fluctuations conditioned on entropy production, we have been led to our central result, which relates the sign of the conditioned entropy production to the direction of time's arrow.


\begin{acknowledgments}
M. B. acknowledges financial support from FAPESP (Brazil), Project No. 2012/07429-0, Unicamp/FAEPEX (Brazil), Grant No. 0031/15 and C. Jarzynski for his hospitality during the visit to University of Maryland, USA.  C. J. acknowledges financial support from the National Science Foundation (USA) under Grant No. DMR-1206971.
\end{acknowledgments}

\appendix*
\section{Results in the long-time limit when $Q$ is defective \label{sec:ap}}

We discuss here how our results are modified when the matrix $Q$ given by Eq.~(\ref{cond10}) is {\it defective} or non-diagonalizable. The main results of sections \ref{sec:cond} and \ref{sec:rever} rely basically on the asymptotic expressions
\begin{subequations}
\begin{align}
\langle \mathbf{1}| Q^{n} | \mathbf{p}(\tau_0) \rangle &\simeq \mu^{n} \langle\mathbf{1}|\mathbf{u}_{1}\rangle\langle\mathbf{v}_{1}|\mathbf{p}(\tau_0)\rangle\,, \label{ap1a}\\
\langle \mathbf{e}_{s}| Q^{n} | \mathbf{p}(\tau_0) \rangle &\simeq \mu^{n} \langle\mathbf{e}_{s}|\mathbf{u}_{1}\rangle\langle\mathbf{v}_{1}|\mathbf{p}(\tau_0)\rangle\,, \\
\langle \mathbf{1} | Q^{n} | \mathbf{e}_{s} \rangle &\simeq \mu^{n} \langle\mathbf{1}|\mathbf{u}_{1}\rangle\langle\mathbf{v}_{1}|\mathbf{e}_{s}\rangle\,,
\end{align}
\label{ap1}
\end{subequations}
which are valid in the limit $n\gg 1$ since $|\alpha_{k}|/\mu < 1$ for $k \geq 2$. In what follows we show how to obtain such asymptotic expressions when $Q$ is non-diagonalizable.

Under the assumptions spelled out in the first two paragraphs of Sec.~\ref{sec:cond} (but dropping the later assumption that $Q$ is diagonalizable) the Perron-Frobenius theorem \citep{seneta} guarantees that the eigenvalue $\alpha_{1} = \mu$ with the largest real part is real and nondegenerate, hence we will continue to write $Q |\mathbf{u}_{1}\rangle = \mu |\mathbf{u}_{1}\rangle$ and $\langle\mathbf{v}_{1}| Q = \mu \langle\mathbf{v}_{1}|$.  However, for those eigenvalues $\alpha_{k}$ whose algebraic and geometric multiplicities differ we have \cite{shilov}
\begin{subequations}
\begin{align}
Q\, |\mathbf{u}_{k}^{(1)}\rangle &= \alpha_{k}\, |\mathbf{u}_{k}^{(1)}\rangle\,, \\
Q\, |\mathbf{u}_{k}^{(\nu_{k})}\rangle &= \alpha_{k}\, |\mathbf{u}_{k}^{(\nu_{k})}\rangle + |\mathbf{u}_{k}^{(\nu_{k}-1)}\rangle
\,,\;\mathrm{for}\;1<\nu_{k}\leq\gamma(k)\,,\label{ap3b}
\end{align}
\label{ap3}
\end{subequations}
and
\begin{subequations}
\begin{align}
\langle \mathbf{v}_{k}^{(\gamma(k))}|\,Q &= \alpha_{k}\, \langle \mathbf{v}_{k}^{(\gamma(k))}|\,, \\
\langle \mathbf{v}_{k}^{(\nu_{k})}|\,Q &= \alpha_{k}\, \langle \mathbf{v}_{k}^{(\nu_{k})}| + \langle\mathbf{v}_{k}^{(\nu_{k}+1)}|\,,
\;\mathrm{for}\;1\leq\nu_{k}<\gamma(k)\,,\label{ap4b}
\end{align}
\label{ap4}
\end{subequations}
where $\gamma(k)$ is the degenerancy of $\alpha_{k}$. Equations (\ref{ap3}) and (\ref{ap4}) show that $|\mathbf{u}_{k}^{(1)}\rangle$ and $\langle \mathbf{v}_{k}^{(\gamma(k))}|$ are genuine eigenvectors. The $|\mathbf{u}_{k}^{(\nu_{k})}\rangle$, for $1<\nu_{k}\leq\gamma(k)$, and $\langle \mathbf{v}_{k}^{(\nu_{k})}|$, for $1\leq\nu_{k}<\gamma(k)$, are the so-called {\it generalized} eigenvectors. Together, genuine and generalized eigenvectors form a basis in which $Q$ assumes the Jordan canonical form \cite{shilov}. Therefore, the following relations apply
\begin{subequations}
\begin{align}
\langle \mathbf{v}_{k}^{(\nu_{k})} | \mathbf{u}_{j}^{(\upsilon_{j})} \rangle &= \delta_{k j}\,\delta_{\nu_{k} \upsilon_{j}}\,, \\
\sum_{k}\sum_{\nu_{k}=1}^{\gamma(k)} |\mathbf{u}_{k}^{(\nu_{k})}\rangle\langle \mathbf{v}_{k}^{(\nu_{k})}| &= \mathds{1}\,,
\label{ap2b}
\end{align}
\label{ap2} 
\end{subequations}
which are the analogue of Eq.~(\ref{cond13b}). 

Equations (\ref{ap3}) imply
\begin{equation}
Q^{n}\,|\mathbf{u}_{k}^{(\nu_{k})}\rangle = \sum_{l=0}^{\nu_{k}-1} \binom{n}{l} \alpha_{k}^{n-l}\, |\mathbf{u}_{k}^{(\nu_{k} - l)}\rangle \,,
\label{ap5}
\end{equation}
for $n \geq \nu_{k}$ and $1 < \nu_{k} \leq \gamma(k)$. Analogously, Eqs.~(\ref{ap4}) imply
\begin{equation}
\langle \mathbf{v}_{k}^{(\nu_{k})}|\,Q^{n} = \sum_{l=0}^{\gamma(k)-\nu_{k}} \binom{n}{l} \alpha_{k}^{n-l}\, \langle \mathbf{v}_{k}^{(\nu_{k} + l)}|\,,
\label{ap6}
\end{equation}
for $n > \gamma(k)-\nu_{k}$ and $1 \leq \nu_{k} < \gamma(k)$.

Expressions (\ref{ap5}) and (\ref{ap6}) allow us to find the dominant contribution of the matrix elements in the left-hand side of Eq.~(\ref{ap1}) when $Q$ is defective. If we take for instance $\langle\mathbf{1}| Q^{n} |\mathbf{p}(\tau_0)\rangle$ and insert Eq.~(\ref{ap2b}), we obtain
\begin{eqnarray}
\langle\mathbf{1}| Q^{n} |\mathbf{p}(\tau_0)\rangle &=& \langle\mathbf{1}| Q^{n}\,\sum_{k}\sum_{\nu_{k}}|\mathbf{u}_{k}^{(\nu_{k})}\rangle\langle\mathbf{v}_{k}^{(\nu_{k})} |\mathbf{p}(\tau_0)\rangle \nonumber \\
&=& \langle\mathbf{1}| Q^{n} |\mathbf{u}_{1}\rangle\langle\mathbf{v}_{1} |\mathbf{p}(\tau_0)\rangle + 
\sum_{k=2}\sum_{\nu_{k}} \langle\mathbf{1}| Q^{n} |\mathbf{u}_{k}^{(\nu_{k})}\rangle\langle\mathbf{v}_{k}^{(\nu_{k})} |\mathbf{p}(\tau_0)\rangle \nonumber \\
&=& \mu^{n} \langle\mathbf{1}|\mathbf{u}_{1}\rangle\langle\mathbf{v}_{1} |\mathbf{p}(\tau_0)\rangle + 
\sum_{k=2}\sum_{\nu_{k}}\sum_{l=0}^{\nu_{k}-1} \binom{n}{l} \alpha_{k}^{n-l} \langle\mathbf{1}| \mathbf{u}_{k}^{(\nu_{k}-l)}\rangle
\langle\mathbf{v}_{k}^{(\nu_{k})} |\mathbf{p}(\tau_0)\rangle\,,
\label{ap7}
\end{eqnarray}
using Eq.~(\ref{ap5}) from the second to the third line. Analogously to Sec.~\ref{sec:cond}, we assume that $\langle\mathbf{1}|\mathbf{u}_{1}\rangle$ and $\langle\mathbf{v}_{1}|\mathbf{p}(\tau_0)\rangle$ are of the same magnitude of $\langle\mathbf{1}|\mathbf{u}_{k}^{(\nu_{k}-l)}\rangle$ and $\langle\mathbf{v}_{k}^{(\nu_{k})}|\mathbf{p}(\tau_0)\rangle$, respectively. Thus, we recover Eq.~(\ref{ap1a}) from Eq.~(\ref{ap7}) if the following limit holds,
\begin{equation}
\lim_{n\to\infty} \binom{n}{l}\frac{\alpha_{k}^{n-l}}{\mu^{n}} = 0\,.
\label{ap8}
\end{equation}

In the limit $n \gg l$, we use an asymptotic expression for the binomial coefficient to obtain
\begin{eqnarray}
\binom{n}{l} \left( \frac{|\alpha_{k}|}{\mu} \right)^{n} &=& \binom{n}{l} \mathrm{e}^{-n \ln{(\mu/|\alpha_{k}|)}} \nonumber \\
&\simeq & \frac{(n/l - 1/2)^{l}\,\mathrm{e}^{l}}{\sqrt{2\pi\,l}}\,\mathrm{e}^{-n \ln{(\mu/|\alpha_{k}|)}} \nonumber \\
& < & n^{l}\,\mathrm{e}^{-n \ln{(\mu/|\alpha_{k}|)}}\,.
\label{ap9}
\end{eqnarray}
Since $\mu/|\alpha_{k}| > 1$ and $l$ is always finite, we obtain
\begin{equation}
\lim_{n\to\infty} n^{l}\,\mathrm{e}^{-n\ln(\mu/|\alpha_{k}|)} = 0\,,
\end{equation}
which, due to Eq.~(\ref{ap9}), implies Eq.~(\ref{ap8}). The same kind of analysis can be done for the other matrix elements of Eq.~(\ref{ap1}). In summary, Eqs.~(\ref{ap1}) still hold and our main results are valid when $Q$ is defective.

\end{document}